\documentclass{optica-article}
\journal{opticajournal}
\articletype{Research Article}
\usepackage{amsmath}
\usepackage{graphicx}
\usepackage{booktabs}
\usepackage{array}
\usepackage{siunitx}
\usepackage{orcidlink}

\newcolumntype{D}{>{$\displaystyle}c<{$}}
\newcommand{\abs}[1]{\left\vert #1 \right\vert}
\renewcommand{\vec}[1]{\boldsymbol{#1}}
\newcommand{\uvec}[1]{\hat{\boldsymbol{#1}}}
\newcommand{\tensor}[1]{\overleftrightarrow{\mathbf{#1}}}
\newcommand{\order}[1]{\mathcal{O}\!\left\{#1\right\}}

\DeclareMathOperator{\tr}{tr}

\begin{document}

\title{Operator-Based Aberration Compensation for Holographic Imaging}

\author{Matthew K. Gronert\,\orcidlink{0000-0002-6552-3027}\authormark{1},
  Anna Prenowitz\,\orcidlink{0009-0001-0521-4898}\authormark{1},
  Kaitlynn Snyder\,\orcidlink{0009-0000-4190-3941}\authormark{1},
  David B. Ruffner\,\orcidlink{0000-0002-8350-1825}\authormark{2},
  Valentina Flores Montes\authormark{2},
  Laura A. Philips\,\orcidlink{0000-0001-6121-4131}\authormark{2},
  and David G. Grier\,\orcidlink{0000-0002-4382-5139}\authormark{1,*},
}

\address{\authormark{1}Department of Physics and Center for Soft Matter Research,
  New York University, New York, NY 10003, USA\\
  \authormark{2}Spheryx, Inc., 330 E. 38th St., New York, New York 10016, USA}

\email{\authormark{*}david.grier@nyu.edu}

\begin{abstract*}
Holograms of micrometer-scale particles encode valuable information about the particles' positions and composition. That information can be extracted with great precision by fitting a recorded hologram to a generative model for the image-formation process. Geometric aberrations introduced by nonideal imaging conditions hamper this analysis by distorting a hologram's recorded intensity pattern. We mitigate this problem by introducing a class of unitary operators whose action on the scattered field reproduces the primary Seidel aberrations. These operators are readily integrated into generative models, including the Lorenz-Mie model for holographic particle characterization. We demonstrate that operator-based aberration compensation restores the accuracy of hologram analysis with minimal calibration or computational overhead. In a real-world demonstration, we show that operator-based aberration compensation effectively eliminates systematic lens-to-lens measurement errors due to normal manufacturing tolerances in commercial objective lenses. The general-purpose operator formalism should have applications in other areas of image analysis and light-field design.
\end{abstract*}

\section{Introduction}
\label{sec:introduction}

\begin{figure}
  \centering
  \includegraphics[width=\textwidth]{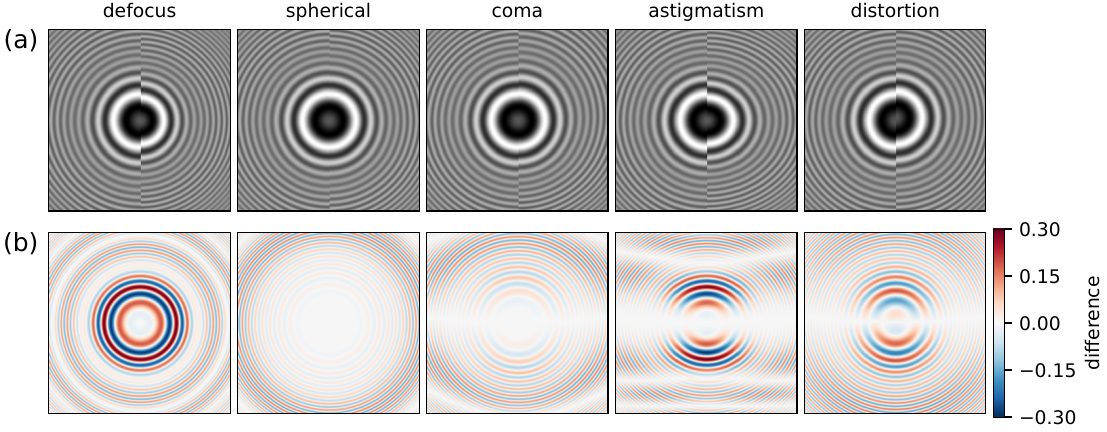}
  \caption{Influence of Seidel aberrations on computed
    holograms of a \qty{1.5}{\um}-diameter
    polystyrene sphere ($n_p = \num{1.614}$)
    in water ($n_m = \num{1.34}$) at
    $z_p = \qty{50}{\um}$ in the middle of the objective's field
    of view. Field-dependent aberrations are centered at
    the edge of the field of view ($\eta = 1$);
    vacuum wavelength: \qty{447}{\nm};
    magnification: \qty{120}{\nm\per pixel}.
    (a) $I(\vec{r}; \vec{\alpha})$ without aberrations, $\alpha_n = 0$
    (left side)
    and with each aberration independently
    (right side),
    with Seidel coefficients set to
    $W_{j\ell m} = 10 \lambda$ for all but distortion
    ($W_{311} = \lambda$),
    for the objectives used in this study.
    (b) Difference, colored relative to mean intensity.}
  \label{fig:gallery}
\end{figure}

When an object scatters coherent light,
information about its position, size, shape and
composition is encoded into the wavefronts of the scattered light.
That information can be retrieved by fitting a
recording of the scattering pattern
to a generative model
for the image-formation process.
Applying such an inverse-problem analysis \cite{lee2007characterizing,soulez2007inverse,martin2022line}
to in-line holograms of micrometer-scale colloidal particles
yields each particle's three-dimensional
position to within a few nanometers, its diameter with nanometer precision, and its refractive
index to within a part per thousand \cite{lee2007characterizing,cheong2011holographic,wang2016holographic,altman2023machine}.
This trove of particle-resolved information
has proved valuable in areas as varied as biopharmaceuticals
\cite{wang2016holographic,kasimbeg2019holographic,winters2020quantitative,chen2023screening}, semiconductor processing \cite{cheong2017holographic,cheong2025detection}
and biomedical research \cite{boltyanskiy2022label,vonhofe2025multivalency,snyder2026simultaneous,quddus2026quantitative}.
All of this value is lost, however, if geometric aberrations, such as the examples in Fig.~\ref{fig:gallery}, distort the recorded holograms
in ways that are not anticipated by the generative model,
so that fits yield spurious results \cite{moyses2013robustness,leahy2020large,martin2021improving,martin2022line,brault2022accurate,snyder2023aberration}.
Such distortions are difficult to avoid in holographic microscopy
because commercial objective lenses are designed and manufactured
to minimize aberrations for objects near the focal plane, not at
the large ranges generally used for holography \cite{leahy2020large}.
Nominally identical
optical trains therefore may distort holograms differently,
causing instrument-to-instrument variations in
downstream quantitative analysis \cite{brault2022accurate}.

The goal of the present work is to mitigate
the effects of aberrations on model-based hologram analysis while retaining the speed
and efficacy of the ideal implementation \cite{grier2026pylorenzmie}.
We do so by modeling geometric aberrations as the action of
unitary operators on the scattered field.
Operator methods have long been used to understand and correct
for Seidel aberrations in the geometric-optics regime,
both for light and for charged particles
\cite{dragt1982lie,hawkes1983lie,forest1987lie,dragt1987elementary,lakshminarayanan1998lie}.
Wolf's phase-space formulation treats geometric optics as
canonical transformations of rays \cite{wolf2004geometric},
and linear canonical transforms have been used to carry Wolf's operator description
to wave fields in the paraxial limit \cite{ozaktas2001fractional}.
A parallel matrix-optics treatment models
how paraxial astigmatism distorts holograms \cite{verrier2014line}.
Aberrations also have been incorporated into holographic
characterization through forward models of the imaging
system
\cite{leahy2020large,martin2021improving}.
This approach is effective, but
requires a detailed
description of the optical train and is computationally costly.
Alternatively, aberrations can be represented by phase
distortions in the pupil plane, either as an expansion in Zernike
polynomials with experimentally calibrated coefficients \cite{brault2022accurate,brault2023multispectral},
or as estimates derived directly from measured holograms
of reference spheres \cite{midtvedt2021fast}.
Zernike calibrations have been shown
\cite{brault2022accurate,brault2023multispectral}
to improve the repeatability of holographic reconstructions,
but require calibration measurements at
multiple positions across the field of view.

Building on the phenomenological operator treatment of Snyder and
Grier \cite{snyder2023aberration}, we use the Lie-operator formulation
of ray propagation developed by Dragt \cite{dragt1982lie} to derive
wave-optics operators for the primary Seidel aberrations.
These operators' action on recordings of scattered fields
can be reduced to closed-form, image-plane phase
corrections that drop into the existing generative model
for Lorenz-Mie holography \cite{lee2007characterizing}
without modeling the full optical train, and at negligible
computational cost.
Because the same framework spans all aberrations,
a small set
of calibration constants obtained by
introspection on a single dataset can mitigate an instrument's
hardware variations for all subsequent measurements.
Using operators to minimize instrument-to-instrument variations maintains
the speed needed for high-throughput applications
and the generality needed for research applications and product development.
We validate operator-based aberration compensation through experiments using both custom-built and commercial
instruments for Total Holographic Characterization
(THC) \cite{lee2007characterizing,martin2022line}.

\section{Operator formalism}

\subsection{Wave-optics description of
aberrations}

A camera records spatial variations
in the time-averaged intensity of a light field:
\begin{equation}
    I(\vec{r})
    =
    \abs{\vec{E}(\vec{r}, t)}^2 .
\end{equation}
Geometric aberrations redistribute the intensity without changing the total flux of light.
Such distortions therefore can be described as the action of unitary operators
$\hat{U}_n(\alpha_n)$ on the fields,
\begin{equation}
    I_n(\vec{r})
    =
    \abs{\hat{U}_n(\alpha_n) \, \vec{E}(\vec{r}, t)}^2 ,
\end{equation}
where $\alpha_n$ is a
parameter describing the severity of the $n$-th aberration.
We obtain expressions for $\hat{U}_n(\alpha_n)$
by noting that
an infinitesimal intensity-preserving distortion can be
expressed as a coordinate transformation,
\begin{equation}
    \vec{E}_n(\vec{r}, t)
    =
    \sqrt{J_n(\vec{r})} \,
    \vec{E}(\vec{r}
    +
    \alpha_n \, \vec{g}_n(\vec{r}), t),
\end{equation}
that is generated by $\vec{g}_n(\vec{r})$,
where
\begin{equation}
    J_n(\vec{r})
    =
    \det\left(
    \tensor{I}
    +
    \alpha_n \nabla \vec{g}_n(\vec{r})\right)
\end{equation}
is the Jacobian of the transformation
and $\tensor{I}$
is the unit tensor.
As shown in Appendix~\ref{app:unitary}, the half-density prefactor,
$\sqrt{J_n(\vec{r})}$, completes the Koopman representation
of the advective flow generated by $\vec{g}_n(\vec{r})$
by conserving the overall flux of light \cite{koopman1931hamiltonian,budisic2012applied}.
We scale $\vec{g}_n(\vec{r})$ so that $\alpha_n$
is measured in units of the vacuum wavelength of light.
Composition of small transformations then yields
the unitary aberration operator,
\begin{equation}
    \hat{U}_n(\alpha_n)
    =
    \exp\left( \alpha_n \, \hat{G}_n \right),
    \label{eq:unitaryoperator}
\end{equation}
whose generator is the
suitably symmetrized, anti-Hermitian
operator,
\begin{equation}
    \hat{G}_n
    =
    \vec{g}_n(\vec{r}) \cdot \nabla
    +
    \frac{1}{2} \left(
    \nabla \cdot \vec{g}_n \right) ,
\end{equation}
that is derived in Appendix~\ref{app:unitary}.
A general distortion can be composed
from a sequence of such operations
\cite{dragt1982lie,lakshminarayanan1998lie},
\begin{equation}
    \hat{U}(\vec{\alpha})
    =
    \prod_{j}
    \hat{U}_{n_j}(\alpha_{n_j}),
\end{equation}
whose ordering matters because the
individual operators do not
commute.
The vector of aberration coefficients,
$\vec{\alpha} = \{\alpha_{n_j}\}$,
serves as a set of instrumental calibration constants.

\subsection{Connecting ray optics to wave optics}
\label{sec:formalism}

\begin{figure}
    \centering
    \includegraphics[width=0.5\textwidth]{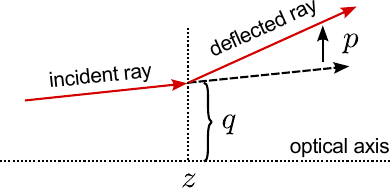}
    \caption{Ray-optics representation of field distortion. A ray propagating along the optical axis is displaced from
    the axis by $\vec{q}(z)$ when it reaches position $z$.
    At that point, the ray is deflected by
    momentum $\vec{p}(\vec{q})$. That deflection corresponds to the imposition of a
    phase gradient at $\vec{R} = (\vec{q}, z)$
    and a redistribution of the field's
    intensity downstream.}
    \label{fig:geometry}
\end{figure}

Expressions for $\vec{g}_n(\vec{r})$ can be obtained
by considering how aberrations deflect rays.
We make this connection through the ray-optics description of the
primary Seidel aberrations \cite{born1999principles} developed
by Dragt \cite{dragt1982lie}.
Figure~\ref{fig:geometry} illustrates
the geometry.
A ray reaching position $z$ along the optical axis
is displaced from the axis by
$\vec{q}(z) = \left(q_x(z), q_y(z) \right)$ in Cartesian coordinates.
The ray has a canonical momentum,
\begin{equation}
  \vec{p}(z)
  =
  n_m(\vec{Q})
  \frac{\partial_z \vec{q}}{
    \sqrt{1 + \abs{\partial_z \vec{q}}^2}
  } ,
\end{equation}
that depends on the refractive index of the medium,
$n_m(\vec{Q})$, at position $\vec{Q} = (\vec{q}, z)$.
With this definition, a ray propagating
through a uniform medium of refractive
index $n_m$ carries momentum
\begin{equation}
    p = \sqrt{p_x^2 + p_y^2} = n_m \, \sin \beta,
\end{equation}
where $\beta$ is the ray's direction of propagation
relative to the optical axis, $\uvec{z}$.
For rays emanating from a scatterer at $\vec{R}_p = (\vec{r}_p, z_p)$,
\begin{equation}
  \sin\beta = \frac{s}{\sqrt{s^2 + z_p^2}},
\end{equation}
where $\vec{s} = \vec{r} - \vec{r}_p$.
The canonical momentum is bounded by the objective's
numerical aperture, $p \leq n_m \sin\beta_\text{max} = \text{NA}$,
which means that the
collected wavefront fills an effective aperture of transverse radius
$\Omega = z_p \tan\beta_\text{max} \approx (\text{NA}/n_m) \, z_p$ in
the imaging plane.
Measuring $r$ relative to $\Omega$ recovers the reduced pupil
coordinate $r/\Omega$ used in
Ref.~\cite{snyder2023aberration}
and clarifies that the distortion caused
by an optical train's
aberrations depends on the axial position of the
object, $z_p$.

By Liouville's theorem \cite{dragt1982lie},
a distribution of rays, $f(\vec{q}, \vec{p})$,
propagates along the optical axis under the influence
of a distortion, $D_n(\vec{q}, \vec{p})$, as
\begin{equation}
  \partial_z f
  =
  \{D_n, f\}
  \equiv
  \sum_{i = 1}^2
  \frac{\partial D_n}{\partial q_i}
  \frac{\partial f}{\partial p_i}
  -
  \frac{\partial D_n}{\partial p_i}
  \frac{\partial f}{\partial q_i} ,
  \label{eq:liouville}
\end{equation}
where $\{\cdot,\cdot\}$ refers to a Poisson bracket.
A distortion therefore displaces a ray according to
\begin{equation}
\label{eq:generator}
    \partial_z \vec{q}
    =
    \{\vec{q}, D_n\}
    =
    \nabla_{\vec{p}} D_n
    \equiv \vec{g}_n(\vec{r}) .
\end{equation}
A defocused lens, for example,
deviates rays according to their momentum \cite{dragt1982lie,born1999principles},
\begin{equation}
    D_\text{defocus}(\vec{q}, \vec{p}) = \abs{\vec{p}}^2,
\end{equation}
so that an arbitrary distribution of rays
propagates as
\begin{equation}
    \partial_z f
    =
    - 2 \vec{p} \cdot \nabla f.
\end{equation}
Applying this to a scattered ray and
comparing with Eq.~\eqref{eq:generator} then
yields the coordinate transformation that describes
how the defocused lens modifies the scattered field:
\begin{equation}
    \vec{g}_\text{defocus}(\vec{r})
    =
    2 n_m \sin\beta \, \uvec{s}.
\end{equation}
This is the first entry in Table~\ref{tab:aberrations}.

The generating functions for other
aberrations in cylindrically symmetric optical systems
follow from their ray-optics
counterparts \cite{dragt1982lie,born1999principles}
in the same manner and are listed in Table~\ref{tab:aberrations}.
Those distortions, $D_n(\vec{q}, \vec{p})$,
that depend on the ray coordinate,
$\vec{q}$, yield generating functions, $\vec{g}_n(\vec{r})$,
that depend on the ray's displacement,
\begin{equation}
  \vec{\eta}(\vec{r}) = \frac{\vec{r} - \vec{r}_c}{H},
\end{equation}
from the center of the aberration, $\vec{r}_c = (x_c, y_c)$,
in the imaging plane.
For applications to microscopy, the displacement is most
conveniently measured relative to the field of view,
$H = \text{FN}/(2M)$, where FN is the objective's field number
and $M$ is its magnification.
The operator formalism is broadly consistent with
Zernike calibrations of holographic microscopes
that find spherical aberration to be uniform across the
field of view \cite{brault2022accurate},
whereas coma and astigmatism vary with position
\cite{brault2022accurate,brault2023multispectral}.

The aberration coefficients, $\vec{\alpha}$,
and the aberration centers, $\vec{r}_c$,
constitute a set of calibration
constants that characterize the instrument's departures from
ideal performance.
The individual coefficients, $\alpha_n$, are related to the standard
Seidel coefficients \cite{welford1986aberrations} by
\begin{equation}
  \alpha_n
  =
  \frac{W_{j\ell m}}{\ell \, \text{NA}^{\ell}},
  \label{eq:seidel}
\end{equation}
where $\text{NA}$ is the numerical aperture of the objective
and $j$, $\ell$ and $m$ are the powers of
$\vec{\eta}$, $\vec{p}$, and $\uvec{s} \cdot \vec{\eta}$,
respectively, in $D_n$ for the $n$-th aberration as listed in Table~\ref{tab:aberrations}.

\begin{table}
  \caption{Generators and phase factors of the Seidel aberrations$^a$}
  \centering
  \renewcommand{\extrarowheight}{6pt}
  \begin{tabular}{c D D D}
    \hline
    \textbf{Aberration}$^b$ &
    D_n(\vec{q}, \vec{p}) &
    \vec{g}_n(\vec{r}) &
    \Phi_n(\vec{r}) \\
    \hline
    defocus ($W_{020}$) &
    \abs{\vec{p}}^2 &
    2 n_m \sin\beta \, \uvec{s} &
    2 n_m k \sin^2\!\beta  \\
    spherical ($W_{040}$) &
    \abs{\vec{p}}^4 &
    4 n_m^3 \sin^3\!\beta \, \uvec{s} &
    4 n_m^3 k \sin^4\!\beta  \\
    coma$^c$ ($W_{131}$) &
    \abs{\vec{p}}^2 \, \vec{p}\cdot \vec{q} &
    n_m^2 \sin^2\!\beta
    \left[2 (\uvec{s} \cdot \vec{\eta}) \, \uvec{s} + \vec{\eta}\right] &
    3 n_m^2 k \sin^3\!\beta \, (\uvec{s} \cdot \vec{\eta})
    \\
    astigmatism ($W_{222}$) &
    \abs{\vec{p}\cdot \vec{q}}^2 &
    2 n_m \sin\beta \, (\uvec{s} \cdot \vec{\eta}) \, \vec{\eta} &
    2 n_m k \sin^2\!\beta \, (\uvec{s} \cdot \vec{\eta})^2
    \\
    field curvature ($W_{220}$) &
    \abs{\vec{p}}^2 \, \abs{\vec{q}}^2 &
    2 n_m \sin\beta \, \abs{\vec{\eta}}^2 \, \uvec{s} &
    2 n_m k \sin^2\!\beta \, \abs{\vec{\eta}}^2
    \\
    distortion ($W_{311}$) &
    \vec{p} \cdot \vec{q} \, \abs{\vec{q}}^2 &
    \abs{\vec{\eta}}^2 \, \vec{\eta} &
    k \sin\beta \, \abs{\vec{\eta}}^2 \, (\uvec{s} \cdot \vec{\eta})
    \\
    \hline
  \end{tabular}
  \label{tab:aberrations}

  \vspace{4pt}
  \raggedright
  \footnotesize
  $^a$Generators of ray deflections, $D_n(\vec{q}, \vec{p})$, from
  \cite{dragt1982lie,born1999principles}
  are associated with generating functions of coordinate transformations,
  $\vec{g}_n(\vec{r})$, and approximate phase distortions,
  $\Phi_n(\vec{r})$, for each aberration.\\
  $^b$Aberrations are associated with the standard Seidel
  coefficients, $W_{j\ell m}$.\\
  $^c$For coma, astigmatism, field curvature and distortion,
  $\vec{\eta}(\vec{r}) = (\vec{r} - \vec{r}_c)/H$
  is the position of the ray in the imaging plane, measured from the
  aberration center, $\vec{r}_c$, in units of the radius of the field of view, $H$.
  The azimuthal angle, $\theta$, is measured relative to $\hat{\vec{x}}$ about the
  particle, so that $\uvec{s} = (\cos\theta, \sin\theta)$ and
  $\vec{p} = n_m \sin\beta \, \uvec{s}$.
\end{table}

\section{Accounting for aberrations in Total Holographic Characterization}
\label{sec:thc}

\subsection{In-line holography}

The operator formalism is useful for
incorporating aberrations into generative models for image formation,
thereby extending their utility
for real-world applications.
As a practical example, we apply
aberration operators to the idealized
generative model used for THC.
The resulting analytical pipeline
efficiently and effectively
corrects for
instrumental imperfections.

The simplest model for in-line hologram
formation assumes that scatterers
are illuminated by a (unit-intensity)
monochromatic plane wave,
\begin{equation}
    \vec{E}_0(\vec{R}, t) = e^{i k z} e^{-i \omega t} \, \hat{x} ,
\end{equation}
at frequency $\omega$ that
propagates along $\hat{z}$ and is
linearly polarized along $\hat{x}$.
The wavenumber $k = 2 \pi n_m/\lambda$
in a medium of refractive index $n_m$ is related to the
frequency by the dispersion relation,
$\omega = c k / n_m$.
Three-dimensional coordinates,
$\vec{R} = (\vec{r}, z)$, are oriented
along the direction of propagation and
the hologram is recorded in the plane $z = 0$.
In the Lorenz-Mie formulation
\cite{bohren2008absorption,mishchenko2002scattering,gouesbet2011generalized},
a particle at $\vec{R}_p = (\vec{r}_p, z_p)$
scatters a proportion of the incident illumination to
the observation point,
\begin{equation}
    \vec{E}_s(\vec{R}, t)
    =
    E_0(\vec{R}_p, t) \,
      \vec{f}_s(k(\vec{R} - \vec{R}_p)) ,
\end{equation}
where $\vec{f}_s(\cdot)$ is the Lorenz-Mie scattering function for $\hat{x}$-polarized
light.
In the particular case of a spherical
scatterer, the Lorenz-Mie function
is parameterized by the sphere's
radius, $a_p$, and its refractive index,
$n_p$, at the imaging wavelength.

The ideal field in the recording plane
is the superposition of $\vec{E}_0(\vec{R}, t)$
and $\vec{E}_s(\vec{R}, t)$, which means that the
generative model for the aberration-free
intensity in the imaging plane is
\begin{subequations}
\begin{align}
    I(\vec{r})
    & =
    \abs{ \vec{E}_0(\vec{R}, t)
    + \vec{E}_s(\vec{R}, t) }^2 \\
    & =
    \abs{ \hat{x} + e^{i k z_p} \, \vec{f}_s(k(\vec{R} - \vec{R}_p)) }^2 .
\end{align}
\label{eq:idealmodel}%
\end{subequations}
This model can be fit to the recorded hologram
of a colloidal particle to estimate the particle's
three-dimensional position, $\vec{R}_p$, as well
as its size and refractive index \cite{lee2007characterizing}.
Highly optimized implementations routinely
yield part-per-thousand precision in each
adjustable parameter in a few milliseconds
\cite{krishnatreya2014measuring,altman2023machine}.
Applying the same model to an aberrated hologram
created by a nonideal imaging system
takes considerably longer and yields
large systematic errors \cite{brault2022accurate,snyder2023aberration}
because Eq.~\eqref{eq:idealmodel}
does not describe the actual image-formation process.

\subsection{Modeling aberrations}

Equation~\eqref{eq:idealmodel} can be generalized to
account for geometric aberrations by applying
unitary operators from Eq.~\eqref{eq:unitaryoperator}
and the generating functions tabulated in Table~\ref{tab:aberrations}.
In this formalism, the aberration operator acts on both the
incident and scattered waves,
\begin{equation}
  I(\vec{r}, \vec{\alpha})
  =
  \abs{\hat{U}(\vec{\alpha})
    \left[\vec{E}_0(\vec{R}, t) + \vec{E}_s(\vec{R}, t) \right]}^2.
  \label{eq:exact}
\end{equation}
The distorted holograms in Fig.~\ref{fig:gallery} were computed
in this way \cite{grier2026pylorenzmie}
using the generating functions
listed in
Table~\ref{tab:aberrations} and the formulation
from Appendix~\ref{app:unitary}.

Applying aberration operators to
the Lorenz-Mie scattering function
incurs a computational cost.
We obtain a fast approximation to Eq.~\eqref{eq:exact}
by considering how $\hat{U}_n(\alpha_n)$
affects the wave scattered by
a Rayleigh particle ($ka_p \ll 1$) \cite{snyder2023aberration,born1999principles},
\begin{equation}
    \vec{f}_R(k\vec{R})
    =
    - i \,
    \frac{n_p^2 - n_m^2}{n_p^2 + 2 n_m^2}
    (ka_p)^3 \,
    \frac{e^{i kR}}{kR} \sin\theta \, \hat{\theta}
\end{equation}
in coordinates centered on the particle, $\vec{R}_p = 0$.
For such scatterers, the principal action of the operator,
\begin{equation}
    \hat{U}_n(\alpha_n) \,
    \vec{f}_R(k \vec{R})
    \approx
    \sqrt{J_n(\vec{r})} \,
    e^{i \alpha_n \Phi_n(\vec{r})} \, \vec{f}_R(k\vec{R})
    +
    \order{\alpha_n
    \frac{\Phi_n(\vec{r})}{kR}}
    \vec{f}_R(k\vec{R}),
    \label{eq:correction}
\end{equation}
is to modify the scattered field with a
multiplicative factor.
Corrections
become small enough to ignore in the far field,
$kR \gg 1$,
which is the case for standard implementations
of THC \cite{lee2007characterizing}.
Applying this eikonal approximation to
Eq.~\eqref{eq:exact} yields
a generalization of the generative model
for the normalized intensity in the microscope's focal plane,
\begin{equation}
  \label{eq:phasemodel}
  I(\vec{r}; \vec{\alpha})
  \approx
  \abs{ \hat{x}
    +
    e^{i \Phi(\vec{r}; \vec{r}_p, \vec{\alpha})} \,
    e^{i k z_p} \vec{f}_s(k(\vec{R} - \vec{R}_p))}^2,
\end{equation}
that compensates for geometric aberrations by modifying
the phase of the scattered field.
The overall phase distortion,
$\Phi(\vec{r}; \vec{r}_p, \vec{\alpha}) = \sum_n \alpha_n \Phi_n(\vec{r})$,
which depends on the particle's position through $\beta$, $\theta$ and $z_p$,
includes contributions from each aberration in the optical
train, although the order of operations no longer matters.
Equation~\eqref{eq:phasemodel} does not
include the
composed density, $J(\vec{r}) = \prod_n J_n(\vec{r})$,
because common factors are divided out when holograms
are normalized for analysis
\cite{lee2007characterizing}.
The phase factors, $\Phi_n(\vec{r})$, associated with the primary
Seidel aberrations are derived in Appendix~\ref{app:eikonal} and are listed
in Table~\ref{tab:aberrations}.
The eikonal approximation is faster to compute than the
exact operator formulation and yields comparable results for
the experimental conditions in this study.

\begin{figure}
  \centering
  \includegraphics[width=\textwidth]{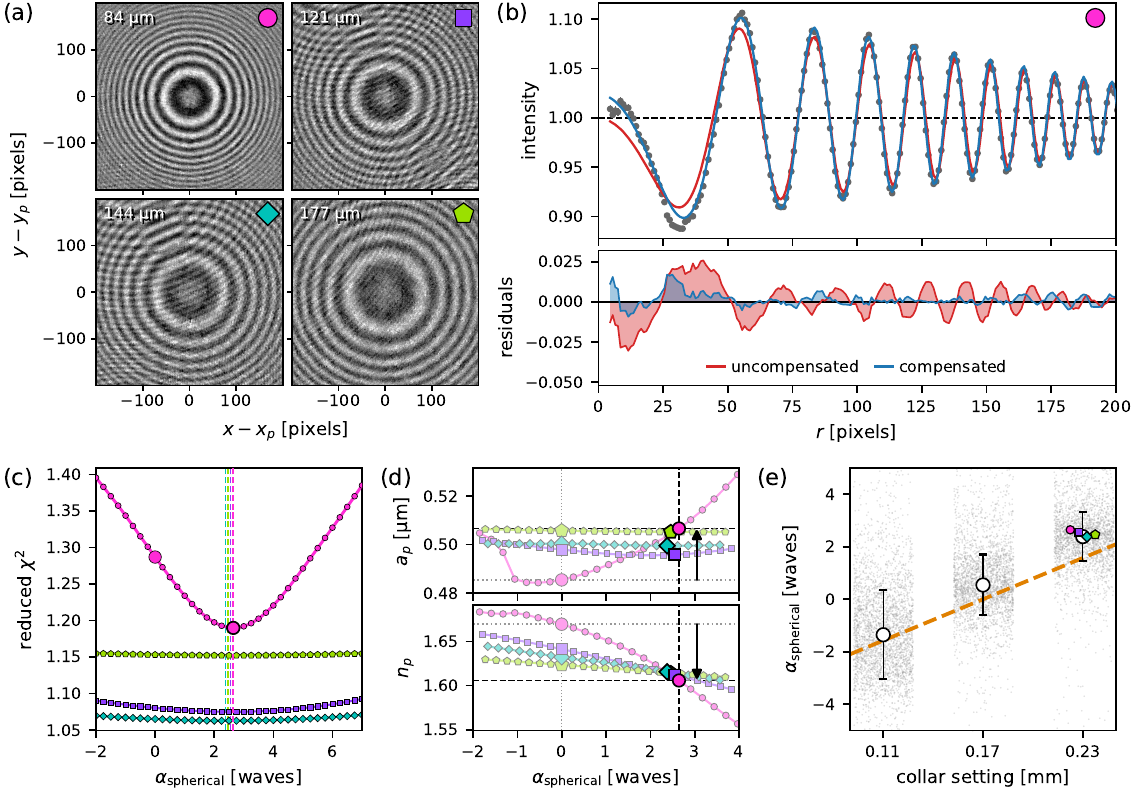}
  \caption{Compensating for spherical aberration.
    (a) Normalized holograms of four polystyrene spheres of
    nominal diameter \qty{1}{\um} in water
    at different axial positions, $z_p$,
    recorded with Lens B.
    (b) Azimuthal average of the hologram at
    $z_p = \qty{84}{\um}$ (points) together with
    best fits to Eq.~\eqref{eq:phasemodel}
    both without (red) and with (blue) compensation
    for spherical aberration.
    Below: residuals.
    (c) Reduced $\chi^2$ for fits with $\alpha_\text{spherical}$
    held fixed for the four particles in (a).
    Dashed lines identify the optimal $\alpha_\text{spherical}$
    for each particle.
    (d) Radius, $a_p$, and refractive index, $n_p$, of the four particles
    obtained at each fixed value of $\alpha_\text{spherical}$.
    Dotted lines mark uncompensated results
    ($\alpha_\text{spherical} = 0$) for the particle at
    $z_p = \qty{84}{\um}$.
    Dashed lines mark values at optimal compensation.
    (e) Median best fit $\alpha_\text{spherical}$ over \num{3000}-particle
    samples of \qty{1}{\um} polystyrene spheres
    for three settings of the objective's
    correction collar.
    Bars are MAD estimates for the standard deviation.
    The dashed line is the theoretical dependence of
    $\alpha_\text{spherical}$ on correction-collar setting,
    Eq.~\eqref{eq:collar}.
    Colored symbols correspond to the particles in (a).}
  \label{fig:correction}
\end{figure}

\subsection{Compensating for aberrations in single-sphere measurements}
\label{sec:single_sphere}

Figure~\ref{fig:correction} demonstrates
the efficacy of
operator-based aberration compensation
for a sample of \qty{1}{\um}-diameter polystyrene
spheres
(Thermo Scientific 5100B)
dispersed in water.
A statistical sample of these particles is
obtained by flowing the dispersion through
the microscope's observation volume in
a microfluidic channel created by bonding
the edges of a \#1.5 glass coverslip to the
face of a glass microscope slide.
The four representative holograms in
Fig.~\ref{fig:correction}(a) were
recorded with a custom-built in-line
holographic microscope \cite{obrien2019above}
operating at a vacuum wavelength of
$\lambda = \qty{447}{\nm}$.
For the present study, the microscope was outfitted with a
Nikon model MRD00405 CFI Plan Apo Lambda 40\texttimes\ air objective
(NA~\num{0.95})
coupled with a \qty{200}{\mm} tube lens
to achieve a system magnification of
\qty{120}{\nm\per pixel} on the sensor of a
video camera (PointGrey, Flea 3, USB 3.0).
We refer to this objective as ``Lens B'' both in
the caption to Fig.~\ref{fig:correction}
and also in Sec.~\ref{sec:compensatingTHC}.
The four holograms differ in scale because each sphere passed through the imaging
volume at a different height, $z_p$,
relative to the microscope's focal plane.
The heights reported in Fig.~\ref{fig:correction}(a) are obtained
by fitting to Eq.~\eqref{eq:phasemodel}.

Holograms were recorded with
the objective's correction collar set to \qty{0.23}{\mm},
which is suitable for a \#2 glass coverslip rather than the
\#1.5 coverslip that was used to assemble
the channel.
Figure~\ref{fig:correction}(b) shows the
subtle yet significant effect of the
spherical aberration introduced by that deliberate
misadjustment.
Without compensation, the best fit (red curve) \cite{grier2026pylorenzmie} of Eq.~\eqref{eq:idealmodel} to the data (discrete points)
systematically misses the positions and amplitudes of the interference fringes.
Correlated errors are clearly visible in the
(red) residual plot.
Optimizing $\alpha_\text{spherical}$ along with
$\vec{r}_p$, $a_p$ and $n_p$
effectively eliminates those residuals (blue curves).

Figure~\ref{fig:correction}(c) shows how the
$\chi^2$ statistic varies when fits are performed
with fixed values of $\alpha_\text{spherical}$.
Such fits converge to a common well-defined
minimum in the $\chi^2$ statistic for all four particles,
as would be expected for an instrumental effect.
Although hologram distortions are most pronounced
for particles closer to the focal plane,
correcting for spherical aberration
substantially influences the
estimated radii, $a_p$, and
refractive indexes, $n_p$, for all four particles, as can be seen in
Fig.~\ref{fig:correction}(d).
Most notably, the phase-only correction
brings all four particles' refractive indexes
into agreement with the value of \num{1.614(1)} for
bulk polystyrene at \qty{447}{\nm} and \qty{20}{\celsius}, which
is indicated by a dashed line
in Fig.~\ref{fig:correction}(d).

Figure~\ref{fig:correction}(e) summarizes
several thousand similar measurements
performed with the objective's correction collar in each of its three settings.
Circular markers identify the median value
of $\alpha_\text{spherical}$ at each setting for \num{3000}-particle statistical samples, with error bars computed as median-absolute-deviation
(MAD) estimates for the standard
deviation.
These results are compared with the expected
value \cite{born1999principles,gibson1991experimental},
\begin{equation}
  \alpha_\text{spherical} =
  \frac{\Delta t}{32 \, \lambda \, n_g}
  \left(1 - \frac{1}{n_g^2} \right),
  \label{eq:collar}
\end{equation}
when the thickness of a coverslip of
refractive index $n_g$ differs from
the collar setting by $\Delta t$.
For a standard coverslip made of
Schott D~263 glass ($n_g = \num{1.534}$ at
$\lambda = \qty{447}{\nm}$),
adjusting the collar should change
$\alpha_\text{spherical}$ by
\num{26} waves per millimeter of collar
setting.
That dependence is plotted as the dashed
line in Fig.~\ref{fig:correction}(e).
Measured and predicted
values for $\alpha_\text{spherical}$ agree for spheres spanning more than \qty{100}{\um} in axial range,
which validates the phase-factor correction.
The eikonal approximation is found to perform comparably
well in measurements on \qty{1}{\um}-diameter silica spheres
and \qty{3}{\um}-diameter colloids composed of either polystyrene or silica
(not shown; see Ref.~\cite{dataavailability}).

\subsection{Compensating for aberrations in THC}
\label{sec:compensatingTHC}

\begin{figure}
  \centering
  \includegraphics[width=\textwidth]{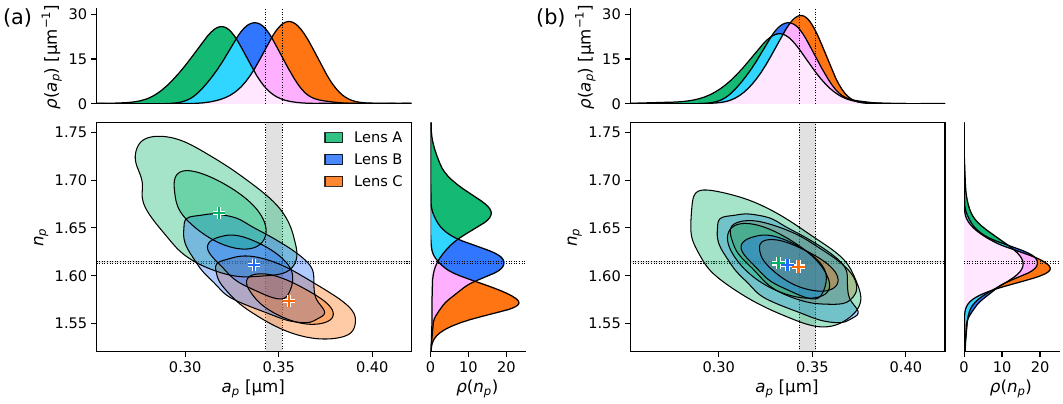}
  \caption{Correcting instrument-to-instrument variations
    in holographic microscopy.
    THC measurements of nominally \qty{700}{\nm} diameter polystyrene
    spheres performed with three nominally interchangeable commercial
    objective lenses.
    Contours enclose \qty{95}{\percent} and \qty{68}{\percent} of
    results in the ($a_p$, $n_p$) plane for \num{1100} particles per lens.
    Projections show size and refractive-index
    distributions.
    Crosses mark the median of each distribution. Gray bars
    span the certified range for radius and refractive index.
    (a) Uncorrected: the three lenses yield inconsistent results.
    (b) With compensation for spherical aberration.}
  \label{fig:monodisperse}
\end{figure}

To illustrate the importance of aberration compensation
for industrial applications of holographic microscopy,
we characterized NIST-traceable polystyrene size standards
(Bangs Labs NT12N, Lot \#11526) with a fully automated
commercial implementation of THC (xSight, Spheryx, Inc.).
The dataset in Fig.~\ref{fig:monodisperse}(a)
emulates instrument-to-instrument variations by exchanging
three nominally interchangeable objective lenses between measurements,
including Lens B from Sec.~\ref{sec:single_sphere}.
The other two (Lens A and Lens C) are
Nikon model MRD70470 CFI Plan Apo Lambda D air objectives,
which also provide 40\texttimes\ magnification at
NA~\num{0.95}.
All three lenses
are installed with correction collars set to \qty{0.22}{\mm},
which is appropriate for the microfluidic sample cells
used in the instrument (xCell50, Spheryx, Inc.).

The lens manufacturer corrects geometric and chromatic
aberrations in each lens for samples near the focal
plane.
Holographic microscopy, however, is based on light scattered
by particles so far from that plane
that standard corrections are unlikely to be effective
\cite{gibson1991experimental,botcherby2007aberration,booth2014adaptive,martin2021improving}.
Figure~\ref{fig:monodisperse}(a) shows that the three lenses
yield markedly different results when their holograms are
analyzed with the ideal generative model from
Eq.~\eqref{eq:idealmodel}.
Population-averaged estimates for the particles'
radius range from $a_p = \qty{319(1)}{\nm}$ (Lens C)
to \qty{357(1)}{\nm} (Lens A).
These values differ significantly from the certified radius
of \qty{347(5)}{\nm}, which the
manufacturer obtained by disc centrifugation.
Results from Lens B are consistent with the certified value.
The magnitude and sign of the discrepancy could not have been anticipated
from measurements on one lens alone.
Similarly, population-averaged estimates for the
refractive index range from
\num{1.5731(4)} (Lens A) to \num{1.662(1)} (Lens C),
and differ significantly from the bulk value of
\num{1.614(1)}.
Adjusting the lenses' correction collars over their full range does not
appreciably improve the lens-to-lens agreement or consistency with
anticipated ground-truth values.

Figure~\ref{fig:monodisperse}(b) confirms that
reanalyzing the same datasets with compensation for spherical
aberration collapses results from all three
lenses onto a single distribution with \qty{92}{\percent} overlap
\cite{cover1967nearest,schilling1986multivariate}.
The optimal values for
$\alpha_\text{spherical}$ are sizable,
ranging from \qty{-6.91}{ waves} for Lens A to
\qty{+4.94}{ waves} for Lens C.
The population-averaged radius now falls between \qty{334(1)}{\nm} and
\qty{343(1)}{\nm}, and the refractive index between \num{1.6084(4)} and
\num{1.6097(6)}.
These differ systematically from the
certified values, but by substantially less than
the uncompensated results.

\begin{figure}
  \centering
  \includegraphics[width=\textwidth]{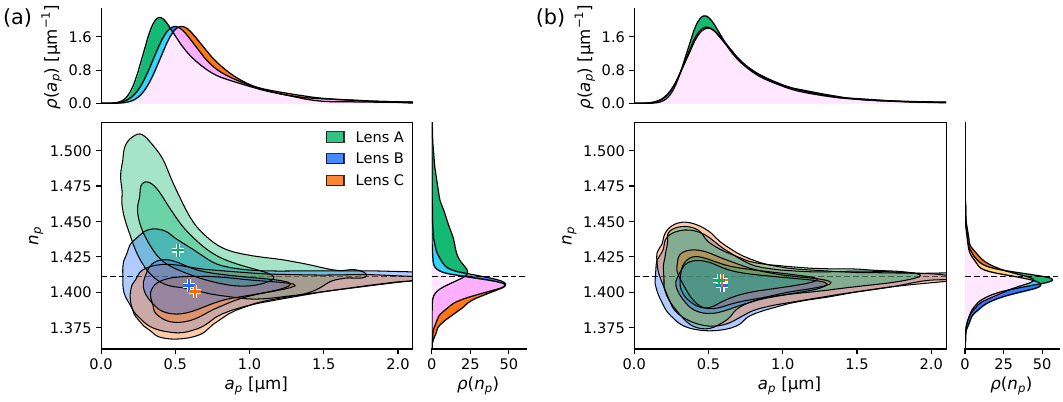}
  \caption{Aberration compensation for a polydisperse emulsion of
    silicone oil droplets using the same three lenses as in
    Fig.~\ref{fig:monodisperse}.
    (a) Uncompensated results disagree,
    especially for the smallest droplets.
    (b) Compensation with $\alpha_\text{spherical}$ from
    Fig.~\ref{fig:monodisperse}(b) yields consistent size and
    refractive-index distributions.
    Crosses mark the median of each distribution.
    The dashed line is the bulk refractive index of the oil.}
  \label{fig:polydisperse}
\end{figure}

Figure~\ref{fig:polydisperse} extends this test to a polydisperse
emulsion of silicone oil droplets in water
(Sigma-Aldrich 378399, Lot \#MKCB2997V).
The droplets span a wide range of sizes but all consist of the same
oil, and so should all share its bulk refractive index,
$n_p = \num{1.411(1)}$, as determined
with an Abbe refractometer
(Edmund Optics \#52-975).
Uncompensated measurements plotted in
Fig.~\ref{fig:polydisperse}(a) yield
inconsistent results for droplet size
and refractive index, especially
for droplets smaller than \qty{1.5}{\um}.
Such errors would hinder material identification based on refractive index
\cite{kasimbeg2019holographic,winters2020quantitative,boltyanskiy2022label}
and distort the size distributions used
to guide manufacturing process control
\cite{markus2024monitoring,chen2026impact}.
Aberration compensation
effectively eliminates size-dependent
systematic errors in Fig.~\ref{fig:polydisperse}(b).
This data collapse, moreover, was achieved
using the same fixed values of $\alpha_\text{spherical}$
obtained for the
monodisperse polystyrene spheres in Fig.~\ref{fig:monodisperse}(b).
This demonstrates that aberration coefficients can be measured
through data introspection on one sample and then used as
calibration constants for subsequent measurements on other samples,
thereby mitigating instrument-to-instrument variations.

\section{Discussion and conclusions}

The operator formalism developed in
this study uses the ray-optics Lie algebra introduced by Dragt \cite{dragt1982lie,dragt1987elementary}
and Wolf's phase-space optics \cite{wolf2004geometric}
to construct a unitary wave-optics description of light propagation
through optical trains.
We have used this formalism to describe how in-line holograms
are distorted by the primary Seidel aberrations.
This framework extends the phenomenological treatment of spherical
aberration by Snyder and Grier \cite{snyder2023aberration}
while placing the analytical technique on a firmer fundamental footing.
The resulting aberration-compensated
analytical pipeline is no more costly
to evaluate than the
standard aberration-free
generative model and
is parameterized
by a small set of aberration coefficients that serve as
instrument-specific calibration constants.
We have demonstrated through applications to THC that
coefficients for aberration compensation can be determined through data introspection
without relying on orthogonal
measurements,
and then can be reused
without requiring per-sample calibration.

The same operator formalism should prove useful in any
application where a generative model describes image
formation.
Single-molecule localization microscopy
(SMLM) uses astigmatism to
encode axial position and therefore
depends on accurate modeling of
the aberrated point-spread function
\cite{huang2008three,pavani2009three}.
Other imaging modalities that already
estimate aberrations during analysis,
including
quantitative phase imaging (QPI), Fourier ptychography and computational
adaptive optics
\cite{park2018quantitative,zheng2013wide,ou2014embedded,adie2012computational,brault2023multispectral}
would benefit from the efficiency and
generality of the operator implementation,
which replaces comprehensive
wavefront mapping with a few
instrumental calibration constants.

The unitarity of the aberration operators
suggests that they also should be useful for
projection systems,
including those used for holographic optical trapping
\cite{grier2003revolution,roichman2006optical}.
In this application, the inverse operator,
$\hat{U}(-\vec{\alpha})$, could precompensate the field associated
with a desired trapping pattern, thereby mitigating aberrations
introduced by the optical train.
Parameterizing field distortions by a small set of aberration coefficients
promises to be faster and easier than mapping phase distortions and
applying global corrections in the hologram plane
\cite{wulff2006aberration,bowman2010slm,cizmar2010situ,liang2018aberration,hill2024optical,otsu2024situ}.

The application we have demonstrated
to holographic microscopy in general and THC in particular
has immediate real-world benefits across
multiple industrial and academic fields.
THC is widely used for product development and
quality assurance in biopharmaceuticals
\cite{wang2016holographic,kasimbeg2019holographic,chen2026impact,gonzalez2026investigating,wong2026optical}
and semiconductor
processing \cite{cheong2017holographic,cheong2025detection},
among other application areas
\cite{philips2017holographic,valentino2023digital,beres2024merging}.
Effective aberration compensation ensures that
commercial applications can rely on THC to reproducibly
detect and characterize the particles in their products
and processes.
The specific demonstration using submicrometer-diameter
polystyrene spheres achieves state-of-the-art reproducibility
for particle-resolved radii and refractive indexes despite
substantial instrument-to-instrument variations,
and recovers the refractive index of polystyrene to within \qty{0.4}{\percent}
\cite{krishnatreya2014measuring,ortiz2023simultaneous,olsen2024dual}.
Self-calibrated aberration compensation also improves the accuracy of holographic
particle tracking \cite{obrien2019above}
and therefore will advance applications of
quantitative holographic microscopy
in soft-matter physics \cite{roichman2008influence,ruffner2012optical,altman2023anomalous}, biophysics \cite{boltyanskiy2022label,vonhofe2025multivalency,quddus2026quantitative,snyder2026simultaneous}
and chemical engineering \cite{cheong2009flow,katz2010applications,middleton2019optimizing,li2021measuring}.

\appendix

\section{Unitary operators}
\label{app:unitary}

An intensity-preserving distortion is a smooth, invertible change of
the transverse coordinate,
$\vec{r} \to \vec{r} + \alpha_n \, \vec{g}_n(\vec{r})$,
generated by the dimensionless vector field
$\vec{g}_n(\vec{r})$.
Represented on the scattered field, the map is
\begin{equation}
    \vec{E}_n(\vec{r}, t)
    =
    \sqrt{J_n(\vec{r})} \,
    \vec{E}(\vec{r} + \alpha_n \, \vec{g}_n(\vec{r}), t) ,
    \label{eq:distortionmap}
\end{equation}
with Jacobian $J_n(\vec{r}) = \det(\tensor{I} + \alpha_n \nabla \vec{g}_n)$.
The half-density prefactor, $\sqrt{J_n(\vec{r})}$, ensures that the
transformation is unitary by preserving the field's norm:
\begin{align}
    \int \abs{\vec{E}_n(\vec{r})}^2 \, d\vec{r}
    & =
    \int J_n(\vec{r}) \,
    \abs{\vec{E}(\vec{r} + \alpha_n \vec{g}_n)}^2 \, d\vec{r} \\
    & =
    \int \abs{\vec{E}(\vec{r}')}^2 \, d\vec{r}' ,
\end{align}
where the change of variables, $\vec{r}' = \vec{r} +
\alpha_n \vec{g}_n(\vec{r})$,
requires $d\vec{r}' = J_n(\vec{r}) \, d\vec{r}$.
Equation~\eqref{eq:distortionmap} is the
Koopman representation of the advective flow of $\vec{g}_n(\vec{r})$
\cite{koopman1931hamiltonian,budisic2012applied}.
It carries
the field along the streamlines of $\vec{g}_n(\vec{r})$
while rescaling its
amplitude to preserve the norm
\cite{dragt1982lie,lakshminarayanan1998lie}.
We identify the generator by
noting that
$J_n(\vec{r}) \approx 1 + \alpha_n \tr(\nabla \vec{g}_n) = 1 + \alpha_n \nabla
\cdot \vec{g}_n$
to first order in small distortions,
$\alpha_n < 1$.
An infinitesimal field transformation then
reduces to
\begin{equation}
    \vec{E}_n(\vec{r})
    \approx
    \vec{E}(\vec{r})
    +
    \alpha_n \left[
    \vec{g}_n \cdot \nabla + \tfrac{1}{2} (\nabla \cdot \vec{g}_n)
    \right] \vec{E}(\vec{r}) ,
\end{equation}
from which we conclude that
$\hat{G}_n = \vec{g}_n \cdot \nabla +
\tfrac{1}{2} (\nabla \cdot \vec{g}_n)$.
Having identified the generator of infinitesimal
distortions, we can compose it into the unitary
aberration operator,
$\hat{U}_n(\alpha_n) =
\exp(\alpha_n \hat{G}_n)$.
Whereas $\alpha_n$ is assumed to be small
in the derivation of $\hat{G}_n$,
its magnitude in $\hat{U}_n(\alpha_n)$
is not constrained.

The unitary aberration
operator implements the flow of
$\vec{g}_n(\vec{r})$ dressed by the half-density,
\begin{equation}
    \hat{U}_n(\alpha_n) \, \vec{E}(\vec{r})
    =
    \sqrt{J_{\alpha_n}(\vec{r})} \;
    \vec{E}\!\left(\vec{T}_{\alpha_n}(\vec{r})\right),
    \label{eq:flowmap}
\end{equation}
where $\vec{T}_{\alpha_n}(\vec{r})$ is the solution of $d\vec{r}/d\alpha_n = \vec{g}_n(\vec{r})$
with $\vec{T}_0(\vec{r}) = \vec{r}$, and
where $J_{\alpha_n}(\vec{r})$ is the Jacobian of $\vec{T}_{\alpha_n}(\vec{r})$.
Applying $\hat{U}_n(\alpha_n)$ is therefore equivalent to warping the
field from the Cartesian coordinates, $\vec{r}$, onto the distorted coordinates, $\vec{T}_{\alpha_n}(\vec{r})$, followed by rescaling by
the half-density,
$\sqrt{J_{\alpha_n}(\vec{r})}$.
The unitary distortion associated with one of the Seidel aberrations
therefore can be
implemented in $\order{N}$ operations on $N$ field samples.
This applies also to a composition of $n$ aberrations, for which the
coordinate transformation,
$\vec{T}(\vec{r}) = (\vec{T}_{\alpha_1} \circ \dots \circ \vec{T}_{\alpha_n})(\vec{r})$, can be precomputed once
$\vec{\alpha}$ is known.

The displacement described by Eq.~\eqref{eq:distortionmap} is a single
Euler step of the full flow.
It differs from $\vec{T}_{\alpha_n}(\vec{r})$
at second order in $\alpha_n$ and therefore does not compose:
$\hat{U}_n(\alpha_1) \, \hat{U}_n(\alpha_2) \neq
\hat{U}_n(\alpha_1 + \alpha_2)$.
The generating functions in Table~\ref{tab:aberrations} are bounded,
so the flow exists for every finite $\alpha_n$.
The expansion parameter is the displacement, $\alpha_n \abs{\vec{g}_n(\vec{r})}$,
relative to the scale, $z_p$, on which $\vec{g}_n(\vec{r})$ varies.
When this ratio is small, the displacement approximates the flow to
within the interpolation error.
Otherwise, the distortion is implemented by integrating the flow
and propagating the Jacobian along each trajectory with
$dJ_{\alpha_n}/d\alpha_n = (\nabla \cdot \vec{g}_n) \, J_{\alpha_n}$.
The full operator can thus be applied exactly, rather than through its
phase approximation, for any applications
where the latter is insufficient.

To establish that the symmetrized
generating operators are anti-Hermitian, we use the advective part, $\vec{g}_n \cdot \nabla$,
to form
an inner product with
arbitrary test fields, $f(\vec{r})$ and $h(\vec{r})$,
and integrate by parts:
\begin{align}
    \left< f, \vec{g}_n \cdot \nabla h \right>
    & =
    \int
    f^*(\vec{r}) \, \vec{g}_n(\vec{r}) \cdot \nabla h \, d\vec{r} \\
    & =
    - \int h(\vec{r}) \, \nabla \cdot \left( f^* \, \vec{g}_n \right) d\vec{r} \\
    & =
    \left<
    \left( - \vec{g}_n \cdot \nabla - \nabla \cdot \vec{g}_n \right) f,
    \, h
    \right> ,
\end{align}
assuming that
$f(\vec{r})$ and $h(\vec{r})$ vanish
at infinity.
Recalling that $\vec{g}_n(\vec{r})$ is real-valued,
the adjoint of the advective term is
$(\vec{g}_n \cdot \nabla)^\dagger =
- \vec{g}_n \cdot \nabla - (\nabla \cdot \vec{g}_n)$,
which shows that the advective term is neither Hermitian
nor anti-Hermitian.
The anti-Hermiticity of the symmetrized operator, $\hat{G}_n = \vec{g}_n \cdot \nabla + \tfrac{1}{2} (\nabla \cdot \vec{g}_n)$, follows:
\begin{equation}
    \hat{G}_n^\dagger
    =
    - \vec{g}_n \cdot \nabla
    - (\nabla \cdot \vec{g}_n)
    + \tfrac{1}{2} (\nabla \cdot \vec{g}_n)
    =
    - \hat{G}_n .
\end{equation}
The unitarity of the exponentiated
distortion operators can be further clarified
by defining the Hermitian operator, $\hat{H}_n \equiv i \hat{G}_n$ and
recognizing that
$\hat{U}_n(\alpha_n) = \exp(-i \alpha_n \hat{H}_n)$ is manifestly
unitary for real $\alpha_n$.
The conservativity of the unitary
distortion operators can be appreciated
by noting that
\begin{equation}
    \frac{d}{d\alpha_n} \lVert \hat{U}_n \vec{E} \rVert^2
    =
    \left<
    \hat{U}_n \vec{E}, \,
    ( \hat{G}_n + \hat{G}_n^\dagger ) \, \hat{U}_n \vec{E}
    \right>
    = 0 ,
\end{equation}
which shows that $\hat{U}_n(\alpha_n)$ redistributes
the recorded intensity
without creating
or destroying light.

\section{Eikonal reduction to phase factors}
\label{app:eikonal}

Applying $\hat{U}_n(\alpha_n)$ explicitly
to the Lorenz-Mie scattering function is costly.
The leading effect of the operation, however, is to multiply the scattered
field by a phase factor.
Working in coordinates centered on the particle, $\vec{R}_p = 0$, the field radiated by a sub-wavelength (Rayleigh) scatterer
carries the rapidly oscillating factor, $\exp(i k S)$, where
$S = \sqrt{s^2 + z_p^2}$ is the distance from a
scatterer at height $z_p$ to a point at transverse position $r$ in the
imaging plane, and $k \Delta R \gg 1$.
Under the coordinate transformation
$\vec{r} \to \vec{r} + \alpha_n \, \vec{g}_n(\vec{r})$,
this carrier acquires the phase
\begin{equation}
  e^{i k S}
  \to
  e^{
    i k \left( S + \alpha_n \, \vec{g}_n \cdot \nabla S \right)
  }
  =
  e^{i k S} \, e^{i \alpha_n \Phi_n} ,
\end{equation}
to leading order in $\alpha_n \vec{g}_n(\vec{r})$,
with the geometric phase factor
\begin{equation}
  \Phi_n(\vec{r})
  =
  k \, \vec{g}_n(\vec{r}) \cdot \nabla S .
  \label{eq:phasefactor}
\end{equation}
The amplitude prefactor, $1/(kS)$, and the polarization structure vary
on the scale of $R$ rather than $1/k$, so their contributions are smaller
than the phase term by a relative factor $\order{1/(kS)} \ll 1$ and may
be dropped.
The half-density, $\sqrt{J_n(\vec{r})}$, multiplies the incident and
scattered waves alike.
It cancels in the particle-free background normalization described in
Sec.~\ref{sec:thc}.
These simplifications yield Eq.~\eqref{eq:phasemodel}.

To evaluate these phase factors for the
Seidel aberrations, we
define $\beta$ to be the polar angle of the scattered ray, so that
$\sin\beta = s / S$.
Similarly, $\theta$ is the azimuthal angle
in the imaging plane, so that
$\uvec{s} = (\cos\theta, \sin\theta)$,
$\nabla S = \sin\beta \, \uvec{s}$ and $\vec{p} = n_m \sin\beta \, \uvec{s}$.
Evaluating Eq.~\eqref{eq:phasefactor} for
the radial generating functions, $\vec{g}_\text{defocus}(\vec{r}) = 2 n_m \sin\beta \,
\uvec{s}$ and $\vec{g}_\text{spherical}(\vec{r}) = 4 n_m^3 \sin^3\!\beta \,
\uvec{s}$, gives
\begin{subequations}
\begin{align}
  \Phi_\text{defocus}(\vec{r})
  & =
    2 n_m k \sin^2\beta \\
  \Phi_\text{spherical}(\vec{r})
  & =
    4 n_m^3 k \sin^4\beta ,
\end{align}
\end{subequations}
which are the first two entries of Table~\ref{tab:aberrations}.

The field-dependent aberrations are treated identically, except that
their generators $D_n(\vec{q}, \vec{p})$ also depend on the ray coordinate.
In considering how aberrations act on
the field in the imaging plane,
we identify the point $\vec{r}$ in the imaging plane
with the ray coordinate, $\vec{q} = \vec{\eta}(\vec{r}) =
(\vec{r} - \vec{r}_c)/H$, measured from the aberration center,
$\vec{r}_c$, in units of the radius of the field of view, $H$.
The corresponding ray momentum in the imaging plane is
$\vec{p} = n_m \sin\beta \, \uvec{s}$
where $\beta$ is the angle at which the ray left the scatterer and
$\uvec{s} = (\cos\theta, \sin\theta)$.
Having established the relationship between
the canonical coordinates
and their conjugate momenta in the imaging plane,
the generating functions follow as
$\vec{g}_n(\vec{r}) = \nabla_{\vec{p}} D_n$:
\begin{subequations}
\begin{align}
    \vec{g}_\text{coma}(\vec{r})
    & =
      2 (\vec{p} \cdot \vec{q}) \, \vec{p} + \abs{\vec{p}}^2 \vec{q}
    =
    n_m^2 \sin^2\!\beta
    \left[ 2 (\uvec{s} \cdot \vec{\eta}) \, \uvec{s} + \vec{\eta} \right] , \\
    \vec{g}_\text{astigmatism}(\vec{r})
    & =
    2 (\vec{p} \cdot \vec{q}) \, \vec{q}
    =
    2 n_m \sin\beta \, (\uvec{s} \cdot \vec{\eta}) \, \vec{\eta} , \\
    \vec{g}_\text{field curvature}(\vec{r})
    & =
    2 \abs{\vec{q}}^2 \, \vec{p}
    =
    2 n_m \sin\beta \, \abs{\vec{\eta}}^2 \, \uvec{s} , \\
    \vec{g}_\text{distortion}(\vec{r})
    & =
    \abs{\vec{q}}^2 \vec{q}
    =
    \abs{\vec{\eta}}^2 \vec{\eta} .
\end{align}
\end{subequations}
When applied to the Rayleigh scatterer with
$\nabla S = \sin\beta \, \uvec{s}$,
Eq.~\eqref{eq:phasefactor} then yields
\begin{subequations}
\begin{align}
  \Phi_\text{coma}(\vec{r})
  & =
    3 n_m^2 k \, \sin^3\beta \, (\uvec{s} \cdot \vec{\eta}) , \\
  \Phi_\text{astigmatism}(\vec{r})
  & =
    2 n_m k \, \sin^2\beta \, (\uvec{s} \cdot \vec{\eta})^2 , \\
  \Phi_\text{field curvature}(\vec{r})
  & =
    2 n_m k \, \sin^2\beta \, \abs{\vec{\eta}}^2 , \\
  \Phi_\text{distortion}(\vec{r})
  & =
    k \, \sin\beta \, \abs{\vec{\eta}}^2 \, (\uvec{s}\cdot\vec{\eta}) .
\end{align}
\end{subequations}
These phase factors are of order $j = 1$, 2, 2 and 3 in $\vec{\eta}$,
respectively,
as the Seidel hierarchy requires.
Measuring the optical path difference, $\ell \alpha_n D_n$,
in units of $\lambda/(2 \pi)$ yields
the phase distortion associated with the corresponding aberration,
$\alpha_n \Phi_n(\vec{r})$.
Writing $D_n$ in terms of $\vec{p}/\text{NA}$ then yields Eq.~\eqref{eq:seidel}.

The experimental results presented in Sec.~\ref{sec:single_sphere}
and Sec.~\ref{sec:compensatingTHC} truncate the Seidel hierarchy
at spherical aberration.
Failing to account for field-dependent aberrations could
introduce systematic errors into THC measurements and
axial tracking results.
Fortunately, the presence of such additional aberrations would
be apparent in fitting results for scatterers over a range
of axial positions, $z_p$.
Field-dependent aberrations include axisymmetric components
that depend on $z_p$ through $\eta$.
These contributions would
influence fit values for $\alpha_\text{spherical}$, which is
sensitive to axisymmetric distortions, and therefore would
introduce an apparent dependence on $z_p$ into
estimates for $\alpha_\text{spherical}$.
The lack of such a trend in the experimental data
therefore serves as a self-consistency
check for the truncated implementation.

\begin{backmatter}

\bmsection{Funding}
National Science Foundation (DMR-2428983, DMR-0922680).

\bmsection{Acknowledgments}
This work was supported by the National Science Foundation
through Award No.~DMR-2428983.
The custom-built holographic microscope
used for this study was constructed as shared instrumentation
through NSF Award No.~DMR-0922680.

\textit{Use of AI tools.}
The authors used the Sonnet 5.5 version of Claude Code, an AI coding assistant from Anthropic,
during preparation of this work.
Claude was used to write and run analysis code,
to check derivations and to verify numerical claims.
The scientific content, the interpretation of results,
and the wording of the manuscript are the authors'.
The authors reviewed and verified all AI-assisted code and text,
validated the analysis software against independent calculations,
and take full responsibility for the content of this article.
No AI tool is listed as an author.

\bmsection{Disclosures}
DGG: Spheryx, Inc. (I). DBR, VFM, LAP: Spheryx, Inc. (E).

\bmsection{Data availability}
Data underlying the results presented in this paper are available in
Ref.~\cite{dataavailability}.
The archive includes a 200-frame excerpt of one of the raw video recordings
together with the code that locates and fits the particles in it.
The complete recordings are available from the authors on reasonable request.

\end{backmatter}


\begin{thebibliography}{10}
\newcommand{\enquote}[1]{``#1''}

\bibitem{lee2007characterizing}
S.-H. Lee, Y.~Roichman, G.-R. Yi, \emph{et~al.}, \enquote{Characterizing and
  tracking single colloidal particles with video holographic microscopy,}
  {\protect\JournalTitle{Opt. Express}} \textbf{15}, 18275--18282 (2007).

\bibitem{soulez2007inverse}
F.~Soulez, L.~Denis, C.~Fournier, \emph{et~al.}, \enquote{Inverse-problem
  approach for particle digital holography: accurate location based on local
  optimization,} {\protect\JournalTitle{J. Opt. Soc. Am. A}} \textbf{24},
  1164--1171 (2007).

\bibitem{martin2022line}
C.~Martin, L.~E. Altman, S.~Rawat, \emph{et~al.}, \enquote{In-line holographic
  microscopy with model-based analysis,} {\protect\JournalTitle{Nat. Rev.
  Methods Primers}} \textbf{2}, 83 (2022).

\bibitem{cheong2011holographic}
F.~C. Cheong, K.~Xiao, D.~J. Pine, and D.~G. Grier, \enquote{Holographic
  characterization of individual colloidal spheres' porosities,}
  {\protect\JournalTitle{Soft Matter}} \textbf{7}, 6816--6819 (2011).

\bibitem{wang2016holographic}
C.~Wang, X.~Zhong, D.~B. Ruffner, \emph{et~al.}, \enquote{Holographic
  characterization of protein aggregates,} {\protect\JournalTitle{J. Pharm.
  Sci.}} \textbf{105}, 1074--1085 (2016).

\bibitem{altman2023machine}
L.~E. Altman and D.~G. Grier, \enquote{Machine learning enables precise
  holographic characterization of colloidal materials in real time,}
  {\protect\JournalTitle{Soft Matter}} \textbf{19}, 3002--3014 (2023).

\bibitem{kasimbeg2019holographic}
P.~N. Kasimbeg, F.~C. Cheong, D.~B. Ruffner, \emph{et~al.},
  \enquote{Holographic characterization of protein aggregates in the presence
  of silicone oil and surfactants,} {\protect\JournalTitle{J. Pharm. Sci.}}
  \textbf{108}, 155--161 (2019).

\bibitem{winters2020quantitative}
A.~Winters, F.~C. Cheong, M.~A. Odete, \emph{et~al.}, \enquote{Quantitative
  differentiation of protein aggregates from other subvisible particles in
  viscous mixtures through holographic characterization,}
  {\protect\JournalTitle{J. Pharm. Sci.}} \textbf{109}, 2405--2412 (2020).

\bibitem{chen2023screening}
W.~Chen, D.~Klemm, K.~Gregoritza, \emph{et~al.}, \enquote{Screening techniques
  for monitoring the sub-visible particle formation of free fatty acids in
  biopharmaceuticals,} {\protect\JournalTitle{Eur. J. Pharm. Biopharm.}}
  \textbf{190}, 242--247 (2023).

\bibitem{cheong2017holographic}
F.~C. Cheong, P.~Kasimbeg, D.~B. Ruffner, \emph{et~al.}, \enquote{Holographic
  characterization of colloidal particles in turbid media,}
  {\protect\JournalTitle{Appl. Phys. Lett.}} \textbf{111}, 153702 (2017).

\bibitem{cheong2025detection}
F.~C. Cheong, J.~Lumer, T.~Markus, \emph{et~al.}, \enquote{Detection and
  characterization of {CMP} pad debris in polishing slurries using total
  holographic characterization,} {\protect\JournalTitle{ECS J. Solid State Sci.
  Technol.}} \textbf{14}, 024003 (2025).

\bibitem{boltyanskiy2022label}
R.~Boltyanskiy, M.~A. Odete, F.~C. Cheong, and L.~A. Philips,
  \enquote{Label-free viability assay using in-line holographic video
  microscopy,} {\protect\JournalTitle{Sci. Rep.}} \textbf{12}, 12746 (2022).

\bibitem{vonhofe2025multivalency}
J.~von Hofe, J.~Abacousnac, M.~Chen, \emph{et~al.}, \enquote{Multivalency
  controls the growth and dynamics of a biomolecular condensate,}
  {\protect\JournalTitle{J. Am. Chem. Soc.}} \textbf{147}, 25242--25253 (2025).

\bibitem{snyder2026simultaneous}
K.~Snyder, A.~D. Hollingsworth, F.~C. Cheong, \emph{et~al.},
  \enquote{Simultaneous holographic molecular binding assays with internal
  calibration standards,} {\protect\JournalTitle{Langmuir}} \textbf{42},
  3155--3163 (2026).

\bibitem{quddus2026quantitative}
R.~Quddus, K.~Kirshenbaum, and D.~G. Grier, \enquote{Quantitative holographic
  agglutination assay for immunoglobulin {A},} {\protect\JournalTitle{Biomed.
  Opt. Express}} \textbf{17}, 1125--1137 (2026).

\bibitem{moyses2013robustness}
H.~W. Moyses, B.~J. Krishnatreya, and D.~G. Grier, \enquote{Robustness of
  {Lorenz}-{Mie} microscopy against defects in illumination,}
  {\protect\JournalTitle{Opt. Express}} \textbf{21}, 5968--5973 (2013).

\bibitem{leahy2020large}
B.~Leahy, R.~Alexander, C.~Martin, \emph{et~al.}, \enquote{Large depth-of-field
  tracking of colloidal spheres in holographic microscopy by modeling the
  objective lens,} {\protect\JournalTitle{Opt. Express}} \textbf{28},
  1061--1075 (2020).

\bibitem{martin2021improving}
C.~Martin, B.~Leahy, and V.~N. Manoharan, \enquote{Improving holographic
  particle characterization by modeling spherical aberration,}
  {\protect\JournalTitle{Opt. Express}} \textbf{29}, 18212--18223 (2021).

\bibitem{brault2022accurate}
D.~Brault, T.~Olivier, F.~Soulez, \emph{et~al.}, \enquote{Accurate unsupervised
  estimation of aberrations in digital holographic microscopy for improved
  quantitative reconstruction,} {\protect\JournalTitle{Opt. Express}}
  \textbf{30}, 38383--38404 (2022).

\bibitem{snyder2023aberration}
K.~Snyder and D.~G. Grier, \enquote{Aberration compensation for enhanced
  holographic particle characterization,} {\protect\JournalTitle{Opt. Express}}
  \textbf{31}, 35200--35207 (2023).

\bibitem{grier2026pylorenzmie}
D.~G. Grier, L.~Altman, F.~C. Cheong, \emph{et~al.}, \enquote{pylorenzmie:
  Python tools for holographic characterization of colloidal particles,}
  Software, version 3.12.0 (2026).
  \url{https://doi.org/10.5281/zenodo.20784723}.

\bibitem{dragt1982lie}
A.~J. Dragt, \enquote{Lie algebraic theory of geometrical optics and optical
  aberrations,} {\protect\JournalTitle{J. Opt. Soc. Am.}} \textbf{72}, 372--379
  (1982).

\bibitem{hawkes1983lie}
P.~W. Hawkes, \enquote{Lie algebraic theory of geometrical optics and optical
  aberrations: a comment,} {\protect\JournalTitle{J. Opt. Soc. Am.}}
  \textbf{73}, 122 (1983).

\bibitem{forest1987lie}
E.~Forest, \enquote{Lie algebraic maps and invariants produced by tracking
  codes,} {\protect\JournalTitle{Part. Accel.}} \textbf{22}, 15--34 (1987).

\bibitem{dragt1987elementary}
A.~J. Dragt, \enquote{Elementary and advanced {Lie} algebraic methods with
  applications to accelerator design, electron microscopes, and light optics,}
  {\protect\JournalTitle{Nucl. Instrum. Methods Phys. Res. A}} \textbf{258},
  339--354 (1987).

\bibitem{lakshminarayanan1998lie}
V.~Lakshminarayanan, R.~Sridhar, and R.~Jagannathan, \enquote{Lie algebraic
  treatment of dioptric power and optical aberrations,}
  {\protect\JournalTitle{J. Opt. Soc. Am. A}} \textbf{15}, 2497--2503 (1998).

\bibitem{wolf2004geometric}
K.~B. Wolf, \emph{Geometric Optics on Phase Space} (Springer, 2004).

\bibitem{ozaktas2001fractional}
H.~M. Ozaktas, Z.~Zalevsky, and M.~A. Kutay, \emph{The Fractional {F}ourier
  Transform with Applications in Optics and Signal Processing} (Wiley, 2001).

\bibitem{verrier2014line}
N.~Verrier, C.~Fournier, L.~M{\'e}{\`e}s, and T.~Fournel, \enquote{In-line
  particle holography with an astigmatic beam: setup self-calibration using an
  ``inverse problems'' approach,} {\protect\JournalTitle{Appl. Opt.}}
  \textbf{53}, G147--G156 (2014).

\bibitem{brault2023multispectral}
D.~Brault, T.~Olivier, N.~Faure, \emph{et~al.}, \enquote{Multispectral in-line
  hologram reconstruction with aberration compensation applied to
  {Gram}-stained bacteria microscopy,} {\protect\JournalTitle{Sci. Rep.}}
  \textbf{13}, 14437 (2023).

\bibitem{midtvedt2021fast}
B.~Midtvedt, E.~Ols{\'e}n, F.~Eklund, \emph{et~al.}, \enquote{Fast and accurate
  nanoparticle characterization using deep-learning-enhanced off-axis
  holography,} {\protect\JournalTitle{ACS Nano}} \textbf{15}, 2240--2250
  (2021).

\bibitem{koopman1931hamiltonian}
B.~O. Koopman, \enquote{Hamiltonian systems and transformation in {H}ilbert
  space,} {\protect\JournalTitle{Proc. Natl. Acad. Sci. USA}} \textbf{17},
  315--318 (1931).

\bibitem{budisic2012applied}
M.~Budi{\v{s}}i{\'c}, R.~Mohr, and I.~Mezi{\'c}, \enquote{Applied
  {K}oopmanism,} {\protect\JournalTitle{Chaos}} \textbf{22}, 047510 (2012).

\bibitem{born1999principles}
M.~Born and E.~Wolf, \emph{Principles of Optics: Electromagnetic Theory of
  Propagation, Interference and Diffraction of Light} (Cambridge University
  Press, 1999), 7th ed.

\bibitem{welford1986aberrations}
W.~T. Welford, \emph{Aberrations of Optical Systems} (Adam Hilger, Bristol,
  1986).

\bibitem{bohren2008absorption}
C.~F. Bohren and D.~R. Huffman, \emph{Absorption and Scattering of Light by
  Small Particles} (John Wiley \& Sons, 2008).

\bibitem{mishchenko2002scattering}
M.~I. Mishchenko, L.~D. Travis, and A.~A. Lacis, \emph{Scattering, Absorption,
  and Emission of Light by Small Particles} (Cambridge University Press, 2002).

\bibitem{gouesbet2011generalized}
G.~Gouesbet and G.~Gr{\'e}han, \emph{Generalized {Lorenz-Mie} Theories},
  vol.~31 (Springer, 2011).

\bibitem{krishnatreya2014measuring}
B.~J. Krishnatreya, A.~Colen-Landy, P.~Hasebe, \emph{et~al.},
  \enquote{Measuring {Boltzmann}'s constant through holographic video
  microscopy of a single colloidal sphere,} {\protect\JournalTitle{Am. J.
  Phys.}} \textbf{82}, 23--31 (2014).

\bibitem{obrien2019above}
M.~J. O'Brien and D.~G. Grier, \enquote{Above and beyond: holographic tracking
  of axial displacements in holographic optical tweezers,}
  {\protect\JournalTitle{Opt. Express}} \textbf{27}, 25375 (2019).

\bibitem{gibson1991experimental}
S.~F. Gibson and F.~Lanni, \enquote{Experimental test of an analytical model of
  aberration in an oil-immersion objective lens used in three-dimensional light
  microscopy,} {\protect\JournalTitle{J. Opt. Soc. Am. A}} \textbf{8},
  1601--1613 (1991).

\bibitem{dataavailability}
M.~K. Gronert, A.~Prenowitz, K.~Snyder, \emph{et~al.}, \enquote{{Data for
  Operator-Based Aberration Compensation for Holographic Imaging},} Zenodo
  (2026). \url{https://doi.org/10.5281/zenodo.23865077}.

\bibitem{botcherby2007aberration}
E.~J. Botcherby, R.~Ju{\v{s}}kaitis, M.~J. Booth, and T.~Wilson,
  \enquote{Aberration-free optical refocusing in high numerical aperture
  microscopy,} {\protect\JournalTitle{Opt. Lett.}} \textbf{32}, 2007--2009
  (2007).

\bibitem{booth2014adaptive}
M.~J. Booth, \enquote{Adaptive optical microscopy: the ongoing quest for a
  perfect image,} {\protect\JournalTitle{Light Sci. Appl.}} \textbf{3}, e165
  (2014).

\bibitem{cover1967nearest}
T.~Cover and P.~Hart, \enquote{Nearest neighbor pattern classification,}
  {\protect\JournalTitle{IEEE Trans. Inf. Theory}} \textbf{13}, 21--27 (1967).

\bibitem{schilling1986multivariate}
M.~F. Schilling, \enquote{Multivariate two-sample tests based on nearest
  neighbors,} {\protect\JournalTitle{J. Am. Stat. Assoc.}} \textbf{81},
  799--806 (1986).

\bibitem{markus2024monitoring}
T.~Markus, J.~Lumer, R.~Stasavage, \emph{et~al.}, \enquote{Monitoring
  polysorbate 80 degradation in protein solutions using total holographic
  characterization,} {\protect\JournalTitle{Int. J. Pharm.}} \textbf{652},
  123843 (2024).

\bibitem{chen2026impact}
W.~Chen, C.~Dimitrakopoulos, N.~S. Oltra, \emph{et~al.}, \enquote{The impact of
  silicone oil on the formation of free fatty acids particles,}
  {\protect\JournalTitle{J. Pharm. Sci.}} \textbf{115}, 104465 (2026).

\bibitem{huang2008three}
B.~Huang, W.~Wang, M.~Bates, and X.~Zhuang, \enquote{Three-dimensional
  super-resolution imaging by stochastic optical reconstruction microscopy,}
  {\protect\JournalTitle{Science}} \textbf{319}, 810--813 (2008).

\bibitem{pavani2009three}
S.~R.~P. Pavani, M.~A. Thompson, J.~S. Biteen, \emph{et~al.},
  \enquote{Three-dimensional, single-molecule fluorescence imaging beyond the
  diffraction limit by using a double-helix point spread function,}
  {\protect\JournalTitle{Proc. Natl. Acad. Sci. USA}} \textbf{106}, 2995--2999
  (2009).

\bibitem{park2018quantitative}
Y.~Park, C.~Depeursinge, and G.~Popescu, \enquote{Quantitative phase imaging in
  biomedicine,} {\protect\JournalTitle{Nat. Photonics}} \textbf{12}, 578--589
  (2018).

\bibitem{zheng2013wide}
G.~Zheng, R.~Horstmeyer, and C.~Yang, \enquote{Wide-field, high-resolution
  {F}ourier ptychographic microscopy,} {\protect\JournalTitle{Nat. Photonics}}
  \textbf{7}, 739--745 (2013).

\bibitem{ou2014embedded}
X.~Ou, G.~Zheng, and C.~Yang, \enquote{Embedded pupil function recovery for
  {F}ourier ptychographic microscopy,} {\protect\JournalTitle{Opt. Express}}
  \textbf{22}, 4960--4972 (2014).

\bibitem{adie2012computational}
S.~G. Adie, B.~W. Graf, A.~Ahmad, \emph{et~al.}, \enquote{Computational
  adaptive optics for broadband optical interferometric tomography of
  biological tissue,} {\protect\JournalTitle{Proc. Natl. Acad. Sci. USA}}
  \textbf{109}, 7175--7180 (2012).

\bibitem{grier2003revolution}
D.~G. Grier, \enquote{A revolution in optical manipulation,}
  {\protect\JournalTitle{Nature}} \textbf{424}, 810--816 (2003).

\bibitem{roichman2006optical}
Y.~Roichman, A.~Waldron, E.~Gardel, and D.~G. Grier, \enquote{Optical traps
  with geometric aberrations,} {\protect\JournalTitle{Appl. Opt.}} \textbf{45},
  3425--3429 (2006).

\bibitem{wulff2006aberration}
K.~Wulff, D.~Cole, M.~B. Reicherter, \emph{et~al.}, \enquote{Aberration
  correction in holographic optical tweezers,} {\protect\JournalTitle{Opt.
  Express}} \textbf{14}, 4169--4174 (2006).

\bibitem{bowman2010slm}
R.~W. Bowman, A.~J. Wright, and M.~J. Padgett, \enquote{An {SLM}-based
  {S}hack--{H}artmann wavefront sensor for aberration correction in optical
  tweezers,} {\protect\JournalTitle{J. Opt.}} \textbf{12}, 124004 (2010).

\bibitem{cizmar2010situ}
T.~{\v C}i{\v z}m{\'a}r, M.~Mazilu, and K.~Dholakia, \enquote{In situ wavefront
  correction and its application to micromanipulation,}
  {\protect\JournalTitle{Nat. Photonics}} \textbf{4}, 388--394 (2010).

\bibitem{liang2018aberration}
Y.~Liang, Y.~Cai, Z.~Wang, \emph{et~al.}, \enquote{Aberration correction in
  holographic optical tweezers using a high-order optical vortex,}
  {\protect\JournalTitle{Appl. Opt.}} \textbf{57}, 3618--3623 (2018).

\bibitem{hill2024optical}
P.~Hill, P.~Lunt, J.~Reiter, \emph{et~al.}, \enquote{Optical phase aberration
  correction with an ultracold quantum gas,} {\protect\JournalTitle{Phys. Rev.
  A}} \textbf{110}, 053308 (2024).

\bibitem{otsu2024situ}
T.~Otsu-Hyodo, Y.~Ohtake, and T.~Ando, \enquote{In-situ aberration correction
  for {Laguerre}-{Gaussian} optical tweezers via optimization of orbit shape,}
  {\protect\JournalTitle{Opt. Express}} \textbf{32}, 30943--30954 (2024).

\bibitem{gonzalez2026investigating}
J.~G. Gonzalez-Calero, H.~Rahn, K.~Wu, \emph{et~al.}, \enquote{Investigating
  adeno-associated virus stabilization at the air-water interface through
  competitive adsorption,} {\protect\JournalTitle{J. Pharm. Sci.}}
  \textbf{115}, 104304 (2026).

\bibitem{wong2026optical}
B.~S. Wong, J.~Ling, Y.~Su, and D.~Fu, \enquote{Optical imaging and
  spectroscopic characterization of subvisible particles in protein
  therapeutics,} {\protect\JournalTitle{Adv. Drug Deliv. Rev.}} \textbf{230},
  115779 (2026).

\bibitem{philips2017holographic}
L.~A. Philips, D.~B. Ruffner, F.~C. Cheong, \emph{et~al.}, \enquote{Holographic
  characterization of contaminants in water: differentiation of suspended
  particles in heterogeneous dispersions,} {\protect\JournalTitle{Water Res.}}
  \textbf{122}, 431--439 (2017).

\bibitem{valentino2023digital}
M.~Valentino, D.~G. Sirico, P.~Memmolo, \emph{et~al.}, \enquote{Digital
  holographic approaches to the detection and characterization of microplastics
  in water environments,} {\protect\JournalTitle{Appl. Opt.}} \textbf{62},
  D104--D118 (2023).

\bibitem{beres2024merging}
N.~D. Beres, J.~Burkart, E.~Graf, \emph{et~al.}, \enquote{Merging holography,
  fluorescence, and machine learning for in situ continuous characterization
  and classification of airborne microplastics,} {\protect\JournalTitle{Atmos.
  Meas. Tech.}} \textbf{17}, 6945--6964 (2024).

\bibitem{ortiz2023simultaneous}
U.~Ortiz-Orru{\~n}o, R.~Quidant, N.~F. van Hulst, \emph{et~al.},
  \enquote{Simultaneous sizing and refractive index analysis of heterogeneous
  nanoparticle suspensions,} {\protect\JournalTitle{ACS Nano}} \textbf{17},
  221--229 (2023).

\bibitem{olsen2024dual}
E.~Ols{\'e}n, B.~Garc{\'\i}a~Rodr{\'\i}guez, F.~Sk{\"a}rberg, \emph{et~al.},
  \enquote{Dual-angle interferometric scattering microscopy for optical
  multiparametric particle characterization,} {\protect\JournalTitle{Nano
  Lett.}} \textbf{24}, 1874--1881 (2024).

\bibitem{roichman2008influence}
Y.~Roichman, B.~Sun, A.~Stolarski, and D.~G. Grier, \enquote{Influence of
  nonconservative optical forces on the dynamics of optically trapped colloidal
  spheres: the fountain of probability,} {\protect\JournalTitle{Phys. Rev.
  Lett.}} \textbf{101}, 128301 (2008).

\bibitem{ruffner2012optical}
D.~B. Ruffner and D.~G. Grier, \enquote{Optical conveyors: a class of active
  tractor beams,} {\protect\JournalTitle{Phys. Rev. Lett.}} \textbf{109},
  163903 (2012).

\bibitem{altman2023anomalous}
L.~E. Altman, A.~D. Hollingsworth, and D.~G. Grier, \enquote{Anomalous tumbling
  of colloidal ellipsoids in {Poiseuille} flows,} {\protect\JournalTitle{Phys.
  Rev. E}} \textbf{108}, 034609 (2023).

\bibitem{cheong2009flow}
F.~C. Cheong, B.~Sun, R.~Dreyfus, \emph{et~al.}, \enquote{Flow visualization
  and flow cytometry with holographic video microscopy,}
  {\protect\JournalTitle{Opt. Express}} \textbf{17}, 13071--13079 (2009).

\bibitem{katz2010applications}
J.~Katz and J.~Sheng, \enquote{Applications of holography in fluid mechanics
  and particle dynamics,} {\protect\JournalTitle{Annu. Rev. Fluid Mech.}}
  \textbf{42}, 531--555 (2010).

\bibitem{middleton2019optimizing}
C.~Middleton, M.~D. Hannel, A.~D. Hollingsworth, \emph{et~al.},
  \enquote{Optimizing the synthesis of monodisperse colloidal spheres using
  holographic particle characterization,} {\protect\JournalTitle{Langmuir}}
  \textbf{35}, 6602--6609 (2019).

\bibitem{li2021measuring}
C.~Li, R.~Panday, X.~Gao, \emph{et~al.}, \enquote{Measuring particle dynamics
  in a fluidized bed using digital in-line holography,}
  {\protect\JournalTitle{Chem. Eng. J.}} \textbf{405}, 126824 (2021).

\end{thebibliography}
\end{document}